\documentclass[apj]{emulateapj} 
\usepackage{natbib}

\newcommand{\oiii}{{[O\,{\sc iii}]}}

\newcommand{\nii}{{[N\,{\sc ii}]}}

\newcommand{\sii}{{[S\,{\sc ii}]}}
\newcommand{\hii}{H\,{\sc ii}\rm}
\newcommand{\hb}{{H$\beta$}}
\newcommand{\ha}{{H$\alpha$}}
\newcommand{\oh}{12\,+\,log(O/H)}

\newcommand{\lam}{$\,\lambda$}

\newcommand{\degree}{\ensuremath{^\circ}}
\def\kms{\,\hbox{km}\,\hbox{s}^{-1}}
\def\Msol{\mathrel{\rm M_{\odot}}}

\def\Msolyr{\mathrel{\rm M_{\odot}yr^{-1}}}
\def\Wm2{\,\hbox{W}\,\hbox{m}^{-2}}
                                                                    
\begin{document}
\title{The A2667 Giant Arc at z=1.03: Evidence for Large-scale Shocks at High Redshift}
\author{T.-T. Yuan\altaffilmark{1},  L. J. Kewley\altaffilmark{1,2}, A. M. Swinbank\altaffilmark{3}, J. Richard\altaffilmark{4}}
\altaffiltext{1}{Institute for Astronomy, University of Hawaii, 2680 Woodlawn Drive, Honolulu, HI 96822}
\altaffiltext{2}{ARC Future Fellow, Research School of Astronomy and Astrophysics, The Australian National University, Cotter Road, Weston Creek, ACT 2611}
\altaffiltext{3}{Institute for Computational Cosmology, Department of Physics, Durham University, South Road, Durham DH1 3LE, UK}
\altaffiltext{4}{CRAL, Observatoire de Lyon, Universit\'e Lyon 1, 9 Avenue Ch. Andr\'e, 69561 Saint Genis Laval Cedex, France}

\begin{abstract}
We present the spatially resolved emission line ratio properties of a $\sim$10$^{10}$ M$_{\odot}$ star-forming galaxy 
at redshift $z\,=\,1.03$. This galaxy is gravitationally lensed as a triple-image giant arc behind the massive lensing cluster Abell 2667.
The main image of the galaxy has magnification factors of 14$\pm$2.1 in flux and $\sim 2 \times 7$ in area, yielding an intrinsic spatial resolution 
of 115$-$405 pc after AO correction with OSIRIS at KECK II.  The $\emph{HST}$ morphology shows a clumpy structure and the \ha\ kinematics indicates
a large velocity dispersion with V$_{max}$${\it \sin{(i)}}$/$\sigma\sim$ 0.73, consistent with high redshift disk galaxies of similar masses.
   From the  \nii/\ha\ line ratios,  we find that the central 350 parsec of the galaxy is dominated by  star formation. 
The  \nii/\ha\ line ratios are higher in the outer-disk than in the central regions.  Most noticeably, we find a blue-shifted region of strong  \nii/\ha\ emission in the outer disk.  
Applying our recent \hii\ region and slow-shock models, we propose that this  elevated \nii/\ha\ ratio  region is contaminated by
a significant fraction of shock excitation due to galactic outflows. Our analysis suggests that shocked regions may mimic flat or inverted metallicity gradients at high redshift.
\end{abstract} 


\section{Introduction}
One of the most challenging aspects of current galaxy formation and evolution theory is the modeling of feedback processes associated with star-formation and/or nuclear activities. Feedback is considered to be the key solution to many observational puzzles such as the quenching of star-formation in massive galaxies \citep[e.g.,][]{Bower06,Dekel09b}, the origin of the mass-metallicity (MZ) relation \citep[e.g.,][]{Tremonti04,Finlator08}, the enrichment of the interstellar and intergalactic medium (ISM, IGM) \citep[e.g.,][and references therein]{Veilleux05}, the flatter central dark matter profile than that predicted in $\Lambda$CDM simulations \citep[e.g.,][]{Flores94,Governato12}, and the problem of the formation of bulge-less galaxies \citep[e.g.,][]{Governato10,Pilkington11}.

Galactic-scale winds and outflows are observed in a wide range of galaxy types that are undergoing significant star formation.
Galactic-scale outflows have been frequently observed in (1) local starburst dwarf galaxies \citep{MartinCL99,Kirby11,vanderwel11},
(2) local ultra-luminous infrared galaxies (ULIRGs) \citep{Rupke02,Rupke05,MartinCL05}, and (3) high-redshift sub-millimeter galaxies (SMGs) \citep{Nesvadba07,Alexander10}.
Galactic winds are prevalent at intermediate and high redshift \citep{Shapley03,Nesvadba08a,SatoT09,Weiner09,MartinCL09,MartinCL12,Rubin10,Rubin11}.

In cosmological hydrodynamical simulations, implementation of a constant wind speed and outflow mass fails to reproduce the galaxy stellar mass functions \citep[e.g.,][]{Crain09}.
A momentum-driven wind feedback model is currently favored by observations and can reproduce the stellar mass function and the observed large-scale metal enrichment \citep{Murray05,Oppenheimer08,Finlator08,Dave11b}.  In the momentum-drive wind model, the efficiency of the winds or mass-loading factor ($\dot{M}_{\rm wind}/SFR$) 
 scales with the declining power of the mass or circular velocity ($V_{c}^{-1}$) of the galaxy ($\dot{M}_{\rm wind}/SFR\propto V_{c}^{-1}$).  However, 
whether this simple scaling relation applies to individual  galaxies of a wide mass range at different redshifts remains to be tested. The physical picture of how exactly the winds interact with and recycle the energy and chemical elements to the ISM and IGM is still lacking.   Indeed, \citet{Hopkins12} have recently developed a more realistic stellar feedback model that includes
 a complex multi-phase structure that depends on the interaction between multiple feedback mechanisms operating on different spatial and temporal scales.

Note that past studies of gas inflows/outflows have mostly been based on slit-spectra of a small number of individual  galaxies \citep{SatoT09,Rubin10a} or average composite spectra from stacked
low signal-to-noise spectra \citep{Weiner09,Rubin10}. Recent studies using KECK/LRIS data have significantly expanded the sample of individual galaxies and prevalent 
outflows have been confirmed for hundreds of individual galaxies at $z\sim1$  \citep{Rubin12, MartinCL12}.

To provide a more concrete observational baseline for the  sophisticated feedback models such as \citet{Hopkins12}, the next leap forward is to look for wind signatures in a spatially resolved manner.  Integral field spectroscopic (IFS) observations  are needed to separate the  various spatial components of the multi-phased ISM \citep{Arribas01,Rich10,Rich11}.  Recent  surveys of wide field of view IFS on local galaxies have dramatically expanded the  quantitative analysis for galactic-wide winds and shocks \citep{Rich10,Rich11,Arribas08,Sharp10}.

IFS studies  have  found widespread shock-excitation associated with galactic outflows in massive local galaxies \citep{Farage10,Sharp10,Rich10,Rich11}.
Shocks are common products of winds and can also be induced by gas accretion and tidal flows in mergers \citep{Veilleux05, Farage10,Genzel11,Soto12b}.
Shock excitation contributes to the ionized state of the hot gas in galaxies, which also depends intimately on the chemical abundance of the ISM. 
To identify the location and fraction of shock excitation across an entire galaxy, it is crucial to disentangle the shock induced ionization and energy dissipation from other mechanisms.  

Narrow line ratio maps can be used to probe the ionization nature of the sources \citep{Veilleux02,Veilleux02w}.  Strong emission line ratios such as the BPT diagram \citep{Baldwin81,Veilleux87,
Allen98, Kewley01,Kewley06} remains one of the most practical methods to probe the ionization mechanisms at both low and high redshift. Combined with the most recent photoionization and shock models,  strong line ratios such as \nii/\ha, \oiii/\hb, and \sii/\ha\, offer a powerful tool to identify  the excitation mechanisms in non-obscured regions.  For example, \citet{Rich10} showed that in NGC 839,  the  \nii/\ha, \oiii/\hb, and \sii/\ha\ ratios are systematically lower near the nucleus, yet high along an outflow bi-cone.  The region with high line ratios is  remarkably well fitted by a shock model with slow velocities typical of galactic winds.

At high redshift,  the star-formation rate (SFR) of galaxies is substantially larger than local star-forming (SF) galaxies \citep{Noeske07b,Noeske07a,Wuyts_S11,Sobral12b}, and galactic
outflows are prevalent and powerful \citep{Erb06e,Swinbank07b,Weiner09,Rubin10,Alexander10,Genzel11}.  It is therefore of paramount importance to observe the masses and energies of high redshift winds 
to constrain the relations between the wind efficiency,  SFR and galaxy masses on a broader scale.  

Because of the observational challenges, outflows and super winds at high redshift have mostly been reported for LBGs, 
SMGs, or in quasar absorption line systems. Spatially resolved spectroscopy for high-$z$ galaxies have been limited to the very 
massive starburst galaxies \citep{Cresci10, Genzel11}. With seeing-limited observations, \citet{Cresci10} reported three $z\sim3$ UV-selected galaxies
with inverted metallicity gradients (i.e., lower metallicity in the nucleus than in the outer-disk). The inverted gradient was explained by primordial cold gas being funneled  to the nucleus. These studies demonstrate the power of IFU studies in revealing the underlying physical processes of the gas accretion at high redshifts.

With the aid of magnification from gravitational lensing and adaptive optics (AO), we can begin to probe the details of fainter and less massive galaxies at high redshift \citep{Nesvadba08a, Swinbank07,Jones10a,Sharon12, Wuyts12}.  The highly magnified multiple images in massive clusters can be magnified by  more than ten,
offering a rare opportunity to obtain spatially resolved observations. This technique has enabled the first high resolution measurements of metallicity gradients at high redshift \citep{Stark08,Jones10b,Yuan11}.

To determine the metallicity gradient evolutions with redshift, we
are undertaking a metallicity gradient survey on lensed galaxies behind galaxy clusters.   
Here we report observations of the $z=1.03$ arc behind the massive cluster Abell 2667.
We find  anomalous spatial line ratio variations indicative of large scale shocks. 
 
Throughout this paper we use a $\Lambda$\,CDM cosmology with $H_0$= 70 km s$^{-1}$
Mpc$^{-1}$, $\Omega_M$=0.30 and $\Omega_\Lambda$=0.70. At z=1.0,  1 arcsec corresponds to 8.1 kpc and a look-back time of 9.3 Gigayear (Gyrs). We use solar oxygen abundance 12 + log(O/H)$_{\odot}$=8.66 \citep{Asplund05}.

\section{Observations, Data Reduction and Analysis}\label{obs}
\subsection{Observations}

\begin{figure}[!ht]
\hspace{-.5cm}
\includegraphics[scale=0.6,angle=0]{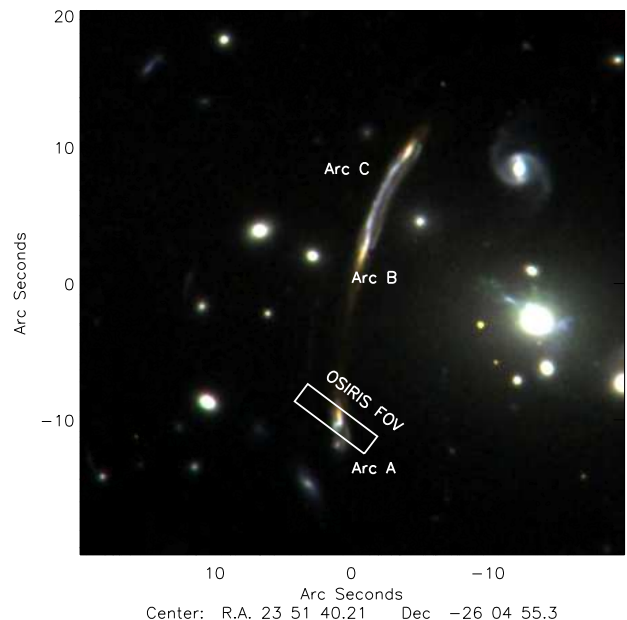}
\vspace{0.8cm}
\caption{Three-color image of the A2667 giant arc at z = 1.03 with $\emph{HST}$ F814W (red), F606W (green), F450W (blue) filters.  
The white box on the bottom left shows the FOV and PA (37\degree) of the OSIRIS Jbb filter.  The triple images composed of arc A, B, C.
The IFU was centered on arc A  $(\alpha_{2000},\delta_{2000})$\,=\,(23:51:40.023, -26:04:51.66). North is up, and East is to the left.}
\label{fig:fov}
\end{figure} 

The A2667 giant arc  is a three-image lensing system. \citet{Covone06} provided the spectroscopic redshift 
of $z = 1.033$ for the arc and the lensing models of the Abell 2667 cluster 
at $z = 0.233$.   We observed the Abell 2667 giant arc with the OH-Suppressing Infra-Red 
Imaging Spectrograph (OSIRIS; \citealt{Larkin06}) in natural guide star Adaptive Optics (NGSAO) mode
on the KECK II telescope at $(\alpha_{2000},\delta_{2000})$\,=\,(23:51:40.023, -26:04:51.66).  
The observation was conducted on the night of September 20-21, 2011, using natural guide star (NGS) AO correction.  The R=14 mag natural guide star was on the edge of the NGS field-of-view (FOV) of OSIRIS, restricting the positional angle (PA) to 37\degree.   The 100 mas pixel scale was used.   Since only broad band filters were available on OSIRIS at the time, we used the Jbb filter (spectral coverage: 1.18 to 1.416$\micron$, spectral resolution: R $\sim$ 3600, FOV
1.6$\arcsec$ $\times$ 6.4$\arcsec$), which covers the \ha, \nii, and \sii\ lines. The observations were conducted in  a standard ABBA dithering sequence,  with the A frame centered on the target and the
B frame centered on an object-free sky region $\sim$20 arcseconds North-East of the target.  The same position angle was used for A and B (sky) frames. 
A total of 5 exposures were obtained for the target,  with 900 seconds on each individual frame. The net on-target exposure time is 75 minutes.   Figure~\ref{fig:fov} shows the OSIRIS FOV and locations of the A2667 arc on the $\emph{HST}$ F814W, F606W, F450W color-coded image.

Before the science target exposure, short exposures of the tip-tilt star (TT) were taken to center the IFU. An offset was applied from the TT star to the science
target.  The offset was determined from astrometry corrected $\emph{HST}$ images with $\sim$ 0.25$\arcsec$ (i.e., $\sim$ 2-3 OSIRIS pixels)  accuracy.  
Gaussian fitting to the point spread function of the TT star yields an average FWHM of 0.13$\arcsec$.  The average FWHM of the sky OH lines is measured to be 
$\sim$ 2.5 $\AA$ for our data and we use this value as the instrumental resolution.  In the analysis below, we deconvolve the line widths with this instrument resolution
by subtracting the instrumental resolution in quadrature from the best-fitting Gaussian $\sigma$.

\subsection{Data Reduction}
Individual exposures were first reduced using the OSIRIS data reduction pipeline \citep{Larkin06}. We used
the sky subtraction IDL code of \citet{Davies07}, which we have modified  to optimize the sky subtraction
specifically around the wavelength region of \ha\ and \nii.  The \sii\ lines are undetected due to  strong telluric absorption. 
Since we are mainly concerned with the emission lines, 
a first order polynomial function was fit to the continuum for each spatial sample pixel (spaxel) and then the continuum was subtracted from each spaxel. The subtraction of the continuum helps to improve the removal of lenslet to lenslet variations. The final datacube was constructed by aligning the sub-exposures with the centroid of the \ha\ images and
combining using a 3$\sigma$ mean clip to reject cosmic rays and bad lenslets. Telluric correction and flux calibration were performed by averaging three A0 standard stars (HIP220, HIP1272, HIP14719)
 observed immediately after the science exposure.  Due to the low declination of this target, the airmass varied between 1.4 to 1.8 during the observation. The uncertainty in absolute flux calibration is estimated to be $\sim$20$-$30\%. However, the relative flux ratio of \nii/\ha\ is accurate to 10\%.

\begin{figure}[!ht]
\hspace{-0.2cm}
\includegraphics[scale=0.5,angle=0]{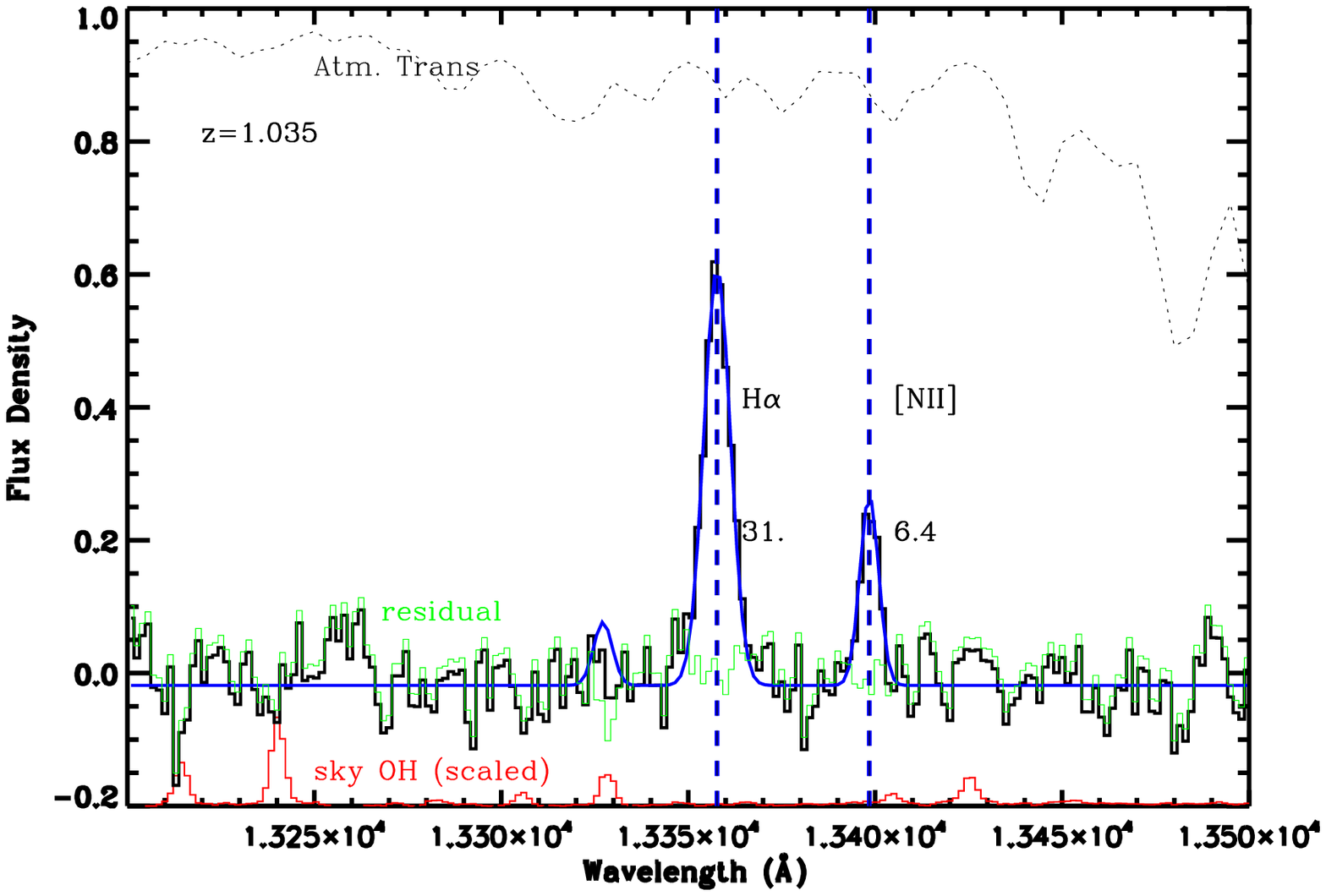}
\caption{Observed spectrum for A2667 arc A, in the vicinity of \ha\ and \nii\ lines.  This is an integrated spectrum  from summing all IFU pixels with \ha\ detections above 5$\sigma$. The flux density is in units of 10$^{-16}$ ergs~s$^{-1}$~cm$^{-2}$~\AA$^{-1}$.   The raw spectrum is shown in black. Gaussian fits to the 3 lines are shown in blue. Note that
the weaker \nii\lam6548 line is fixed to be the theoretical value (1/3) of the \nii\lam6584 line.  The velocity width of \nii\  is constrained to be the same as \ha. Residuals are plotted in green. The locations of the strong sky OH emission (scaled and extracted from a pure sky frame) are shown in red at the bottom of the spectrum.    The atmosphere transmission (black dotted line) is shown at top. The location of \ha\ and \nii\ lines are not contaminated by strong OH lines nor strong telluric absorptions.   
 We achieve a S/N of 37 on \ha\ and of 6.4 on \nii.}
\label{fig:spec}
\end{figure}

\subsection{Analysis}
Our analysis follows the procedures outlined in \citet{Jones10a}. 
Gaussian profiles were fitted simultaneously to the three emission lines:  \nii\lam6548, 6583 and \ha. 
The centroid and velocity width of \nii\lam6548, 6583 lines were constrained by the velocity width of \ha\lam6563, and 
the ratio of \nii\lam6548 and \nii\lam6583 is constrained  to be the theoretical value of  \citet{Osterbrock89}. 
We fit the line profiles  using a weighted $\chi^2$ minimization procedure which takes into account the greater noise level close to atmospheric OH emission.
A blank region of sky within each reduced datacube was used to determine the variance spectrum. Since the variance spectrum is dominated by sky emission,  the fitted spectra were
weighted by the inverse of the variance spectrum (dominated by sky counts) so that regions of strong sky lines do not cause spurious fits.   Fortunately, the sky emission is weak over the \ha, \nii\ wavelength
range for A2667 arcA (Figure~\ref{fig:spec}).  We compute the $\chi^2$ statistic for the best-fitting Gaussian lines and compare with that of a featureless spectrum. 
 We first  fit the spectrum for each pixel, requiring a minimum of $5\sigma$ for detection.  Where this criterion is not met, we average the surrounding 3$\times$3 spatial pixels to achieve higher signal to noise. No fit is made if the 3$\times$3 averaging fails to produce the minimum S/N.   Figure~\ref{fig:spec} shows the integrated spectrum from the image plane. 

In order to check that our results are insensitive to the lensing model uncertainty,  we first analyze the data on the image plane, and then repeat the analysis on the source plane. 
We find that the results are consistent for the image-plane and source-plane analysis.  In the following sections, we will show the spectra measured on the image-plane with associated physical scales 
determined from the source plane.

\section{Global Properties of the Giant Arc}
We determined the total photometry for arc A in 7 bands from $\emph{HST}$ and Spitzer archival data:  $\emph{HST}$ ACS/F814, WFPC2/F606W, WFPC2/F450W, and Spitzer IRAC 3.6, 4.5, 5.8 and 8.0$\mu$m.  Chandra data are also available for the cluster, but we find no X-ray detection on the giant arc.  We used a fixed aperture of diameter 4$\arcsec$ defined in the ACS/F814 band to measure the photometry. 
 For the IRAC colors, we measured the PSF near arc A using the bright isolated stars in the field of view and convolve the ACS/F814 aperture with
this PSF to calculate the aperture flux.     Note that  the photometric aperture covers the entire image of arc A whereas 
 the OSIRIS FOV does not cover the tip of the southern clump C3 (Figure~\ref{fig:fov} for OSIRIS FOV; C3 is labeled on Figure 3). This  southern clump  only contributes to $\sim$ 5\% to the total photometry, well below the $\sim$10\% photometric measurement error. The intrinsic photometric error is dominated by the lensing magnification uncertainty ($\sim$15\%). 
 
We use  the software \verb+LE PHARE+ \citep{Ilbert09} to determine the stellar mass. 
\verb+LE PHARE+ is a photometric redshift and simulation package based on
population synthesis models of \citet{BC03}.  We choose the initial mass function (IMF) by
\citet{Chabrier03} and the \citet{Calzetti00} attenuation law, with E$\rm{(B-V)}$ ranging from 0 to 2 and an exponentially decreasing SFR 
(SFR $\varpropto$ e$^{-t/\tau}$) with $\tau$ varying between 0 and 13 Gyrs.    After correcting for the flux magnification of $\mu=$14$\pm$2.1,
the stellar mass from SED fitting is M$_{\rm{star}}$=$1.90^{+2.0}_{-1.0} \times 10^{10} M_{\odot}$ (lensing uncertainty included), 
assuming the best fit extinction value E$\rm{(B-V)}$=0.6, and SFR$_{\rm{SED}}$ = 50$\pm$35 M$_{\odot}$ yr$^{-1}$. 
Taking E$\rm{(B-V)}_{\rm{gas}}$\,=\, E$\rm{(B-V)}_{\rm{star}}$ / 0.44 \citep{Calzetti97a} and Rv=3.1, we obtain 
A(\ha)=3.46 according to the \citet{Cardelli89} classical extinction curve.  Note that the A(\ha) is $\sim$ 2.46 higher than the median A(\ha) ( $\sim1.0$) at similar masses
from the local and  $z\sim1.7$ star forming galaxy dust extinction vs. stellar mass relation \citep{Garn10,Sobral12}. 

The total \ha\ flux from our OSIRIS spectra is 6.3$\pm$0.2 $\times$10$^{-16}$ ergs~s$^{-1}$~cm$^{-2}$ (before correcting for lensing magnification).
Applying the \citet{Kennicutt98a} SFR prescription and after correcting for lensing magnification and extinction (A(\ha)=3.46), the 
\ha\ star formation rate is estimated to be SFR$_{H\alpha}$\,=\,49.2$\pm$16.5 M$_{\odot}$ yr$^{-1}$. Note that the error is calculated by assuming 
a 15\% uncertainty in lensing magnification and 30\% error in \ha\ flux measurement.  The error estimation excludes the uncertainty in extinction and  aperture flux loss due  to the coverage of the OSIRIS instrument. The physical properties of Arc A is summarized in Table 1.

\section{Lensing Reconstruction, morphology and kinematics}\label{lensing}
\begin{figure}[!ht]
\hspace{-0.4cm}
\includegraphics[scale=0.4,angle=0]{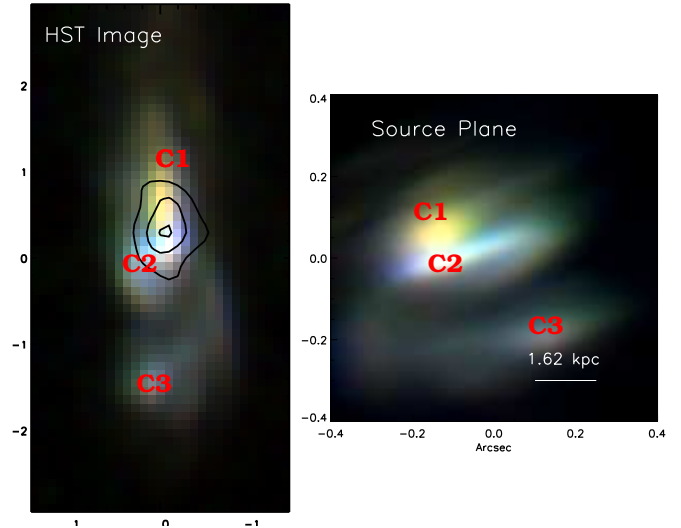}
\caption{Left:  $\emph{HST}$  F814W, F606W, F450W three-color image for A2667 arc A. 
The black contours show the observed \ha\ flux levels of 2, 3, 3.5 in units of 10$^{-17}$ ergs~s$^{-1}$~cm$^{-2}$~arcsec$^{-2}$.
Right: reconstructed source plane morphology from the three-color $\emph{HST}$ image on the left.  The physical scale is marked. 
The distance between the center of the red and blue clumps is $\sim$ 600 pc. 
There are three main clumps revealed in both the observed $\emph{HST}$  and reconstructed image: a northern red component (marked as C1),  and two blue
southern clumps (marked as C2, C3).  The morphology implies either a face-on clumpy disk or a merging system.  North is up and East is to the right. 
}
\label{fig:recons}
\end{figure} 

\subsection{Lens Modeling}
We use the best fit lensing model parameters from \citet{Covone06, Richard10b} and use the software {\tt Lenstool}\footnote{\tt
  http://www.oamp.fr/cosmology/lenstool} \citep{Kneib93, Jullo07} to derive the geometrical transformation between image plane coordinates and source plane coordinates.
The $\emph{HST}$ image and IFU datacube are reconstructed using this mapping and based on the conservation of surface brightness.
The total magnification is computed by the ratio of the sizes (or equivalently, the total fluxes) between the image and its source plane reconstruction.
 As the magnification is not isotropic, the angular size of each image is more highly stretched along a specific axis. However, within the single image  of arc A, the distortions among each pixel is identical.
 Figure~\ref{fig:recons} shows the reconstructed image from the $\emph{HST}$ F814W, F606W, F450W three-color image.   
 The total flux magnification for arc A is $\mu$=14.0$\pm$2.1, and linear magnifications in two axes are $\mu1$=7.0$\pm$1.4, and $\mu2$=2.0$\pm$0.4.

\subsection{Morphology \& Kinematics}

\begin{figure*}[!ht]
\hspace{0.8cm}
\includegraphics[scale=0.75,angle=90]{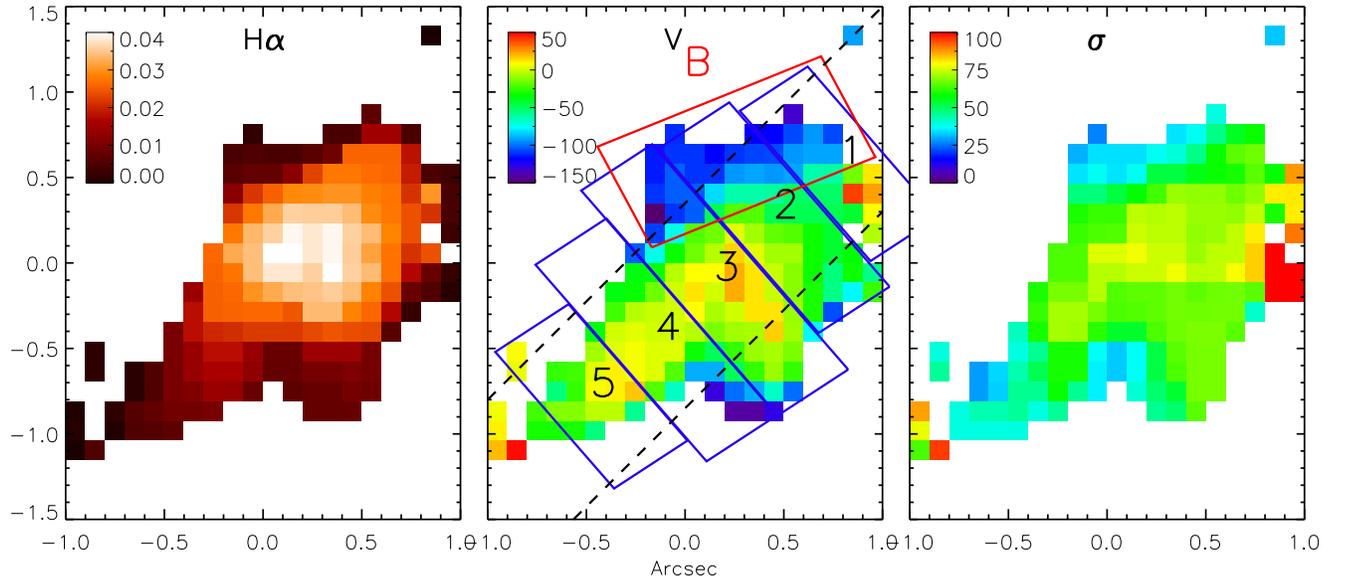}
\vspace{-2.5cm}
\caption{From left to right:  the \ha\ intensity, \ha\ Gaussian fit velocities and velocity dispersions presented on the image-plane.
The \ha\ flux is in units  of 10$^{-17}$ ergs~s$^{-1}$~cm$^{-2}$, and the velocity is in unit of km~s$^{-1}$.
For the middle panel: the black dashed lines indicate the ``slit" used for extracting 1D spectrum in Figure 5;  the blue and red boxes are 
used in the analysis of Section 5.2. These images are shown in  IFS datacube coordinates. North and East are 37\degree\ clockwise from the
$\emph{HST}$ images in Figures 1 and 3. 
}
\label{fig:ha}
\end{figure*} 

\begin{figure}[!ht]
\vspace{-0.3cm}
\hspace{-0.8cm}
\includegraphics[scale=0.4,angle=90]{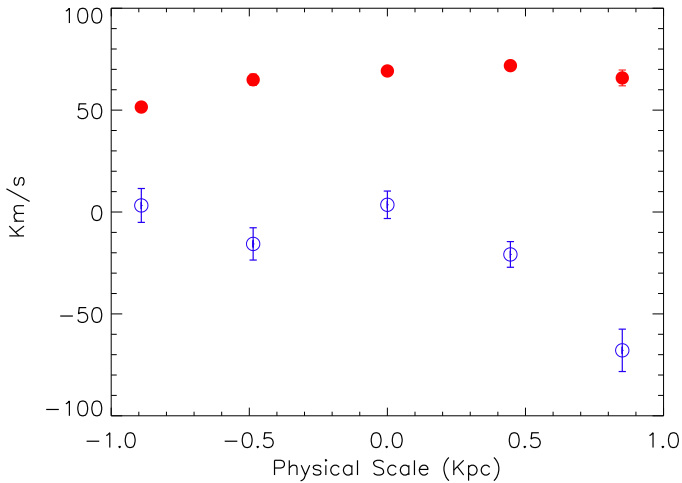}
\caption{ One-dimensional velocity extracted from the slit position shown Figure 4.  The five data bins from left to right correspond
to the direction from box 5 to box 1 in Figure 4. The physical scale is calculated  on the source-plane.
 Note that the errors bars on each data point are the bootstrapped median errors for all the pixels in the bins, weighted by the \ha\ intensity.  
The high velocity dispersion is consistent with those of  turbulent high redshift galaxy disks.}
\label{fig:vel}
\end{figure} 

\begin{figure*}[!ht]
\vspace{-1.5cm}
\begin{center}
\includegraphics[trim = 42mm 72mm 56mm 62mm, clip, width=12.5cm,angle=0]{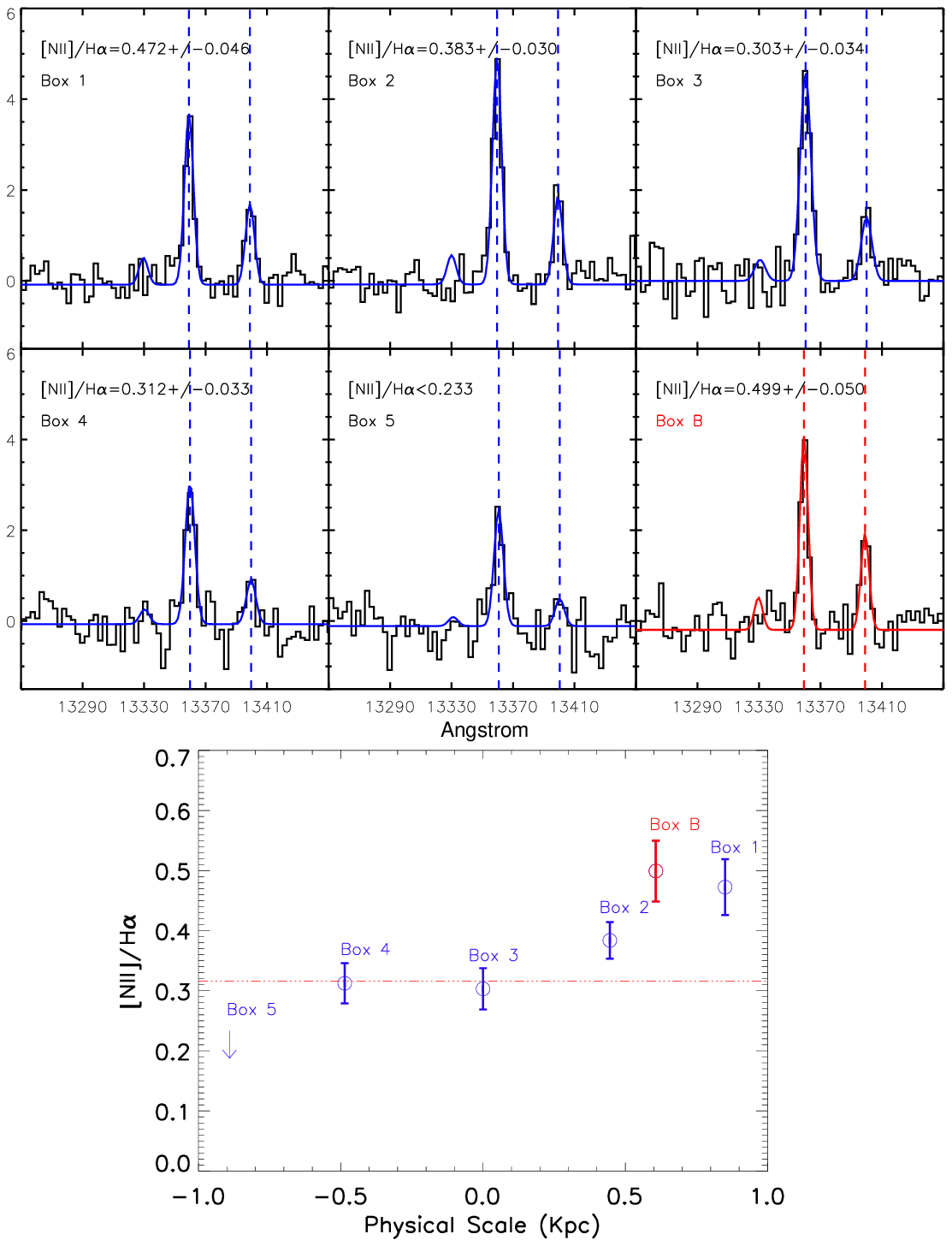}
\caption{The \nii\,/\ha\  ratio variation in boxed regions along the major axis. The top 6 panels  show the observed image-plane spectra in
vicinity of 
 the \nii\ and \ha\  for the 6 boxed regions in Figure 4. The spectra are the \ha\ flux weighted median of the pixels in each box.  The x axes are wavelengths  in \AA, and the y axes are  flux density are in 0.001$\times$10$^{-17}$ ergs~s$^{-1}$~cm$^{-1}$~\AA$^{-1}$.   The \nii\ and \ha\ line positions are marked as vertical dashed lines. The bottom panel shows the \nii\,/\ha\ ratio as a function of the intrinsic physical scale (after lensing correction) of the galaxy.  The red horizontal line indicates the  line ratio corresponding to the solar metallicity value. Ratios above solar can not be directly related to metallicity. We also notice a significant enhancement of \nii\,/\ha\ in Box B, a blueshifted region with respect to the galactic center. }
\label{fig:box}
\end{center}
\vspace{0.2cm}
\end{figure*} 

The $\emph{HST}$ three-color image reveals at least three main clumps (labeled as C1, C2, C3) for arc A. 
These three clumps are easily identified in both the image-plane (Figure~\ref{fig:recons}, Left) and source-plane (Figure~\ref{fig:recons}, Right) morphology. 
The northern red clump (C1) could be dominated by an old stellar population. Alternatively, C1 could be heavily reddened by dust.
After masking out the  two blue clumps (C2, C3) on the south, a Sersic profile fit to the red component on the source plane 
yields a Sersic index of 0.63 $\pm$ 0.30, and a half light radius of $\rm{Re}$\,=\,1.1$\pm$0.2 kpc. The blue clumps (C2, C3) are likely to be active star forming regions. 
The peak of the \ha\ emission is close to the largest blue clump as shown in the \ha\ contours of Figure~\ref{fig:recons}.
One the one hand, the asymmetric clumpy structure of A2667arc is similar to the asymmetric clumpy feature seen in $z\sim2$ disks \citep{Genzel06, Genzel11,Law09,Lehnert09,FS09,FS11}. 
On the other hand, the morphology could also be interpreted as a merging system. We combine the morphology with the velocity structure
to further analyze the disk-versus-merger origin of the  system.

In Figure~\ref{fig:ha}, we show the two-dimensional (2D)  \ha\ intensity, \ha\ Gaussian fit velocity and velocity dispersion maps.
In Figure~\ref{fig:vel}, we extract the  one-dimensional (1D) velocities and dispersions  from a slit oriented along the direction of the blue and red components (black dotted lines
in Figure 4, middle panel).
We divide the spaxels into five bins equally along the slit direction and extract the \ha\ flux weighted spectrum for each bin.  
There is a small velocity gradient ($\sim$75~km~s$^{-1}$) across a 2 kpc length. Since the shape of the 1D velocity curve
is not clearly of a rotating disk, the small velocity gradient could result from  either an almost face-on rotating field or
a non-rotating system.   The system however is clearly dominated by  dispersion with  $\sigma_{median}$/$V_{max} \sin{i}$$\sim$1.3, 
consistent with the large velocity dispersion of the high redshift turbulent disks \citep{Lehnert09,FS09,Jones10a}. 

From our morphology and velocity maps, it remains unclear  whether the A2667 arc  is a face-on clumpy disk or a merging system, especially 
considering that  the \ha\ lines are not observed across the entire galaxy.  Higher S/N data with better spatial coverage is required to more reliably model the velocity field.

\section{Spatially Resolved Line Ratios}
\subsection{\nii\lam6583/\ha\ as an indicator for metallicity and nonphotoionized emission}
The  \nii\lam6583/\ha\ ratio is often used as an indicator for nebular metallicity at high redshift (up to $z\sim2.5$).  
The adjacency of the two lines in wavelength means the line ratio is insensitive to reddening correction and flux calibration. 
The calibration of  \citet{Pettini04} (PP04) is commonly used to convert  \nii\,/\ha\ to metallicity.  
Using the  oxygen abundance of local \hii\ regions, PP04 measured metallicities from the $Te$ (electron temperature) method and determined the correlation
between metallicity and the N2 ($\equiv$ log(\nii\lam6583/\ha)) values as follows:
\begin{equation}
12+{\rm log (O/H)}=8.90+0.57\times N2 
\end{equation}

Note that at high abundance, PP04 used the strong line metallicities rather than the  $Te$ metallicities because of the weakness of the $Te$ sensitive \oiii\lam4363 line. 
Strictly speaking the PP04 relation only holds true for the range of $-1.8\leq N2 \leq-0.8$ (or  $0.0158 \leq$ \nii\,/\ha\ $\leq 0.158$). At extremely low N2 values, the calibration of N2
is difficult due to the weakness of the \nii\ line.  At $N2>-0.8$, the relation is nonlinear. 
The reason for this is, first, as the metallicity becomes larger and  $Te$ decreases, the production of secondary nitrogen begins to dominate and maintain a high value of N2. At even higher metallicity,  the rise in metallicity cools the nebula, lowering $Te$ and allowing less collisional excitations of the 
  \nii\ lines \citep{Kewley02}. We note that the \nii\,/\ha\  ratio is particularly  sensitive to non-thermal excitation.   The presence of shock excitation or an AGN will  increase the \nii\,/\ha\ ratio  and cause abundances determined to be artificially high \citep{Baldwin81, Kewley02, Rich10}.  For metallicity studies,  it is crucial to check if the N2 values are within the regime of pure star formation.

On the other hand, if N2 surpasses the linear metallicity diagnostic range,  the high N2 ratios may be a smoking gun for the presence of non-photoionization mechanisms  such as shocks.   For example,  \citet{Veilleux02w} studied the \nii\,/\ha\ distribution in the local super-wind galaxy M82. They found that  for \nii\,/\ha~$>0.5$, additional excitation mechanisms such as shocks
are needed to explain the enhanced  \nii\,/\ha\  ratio.  Using the most recent shock and photoionization models with IFU observations,  \citet{Rich11}  showed that for the local galaxy IC 1623, shocks contribute  more than 40\% of the line ratio of  \nii\,/\ha.

\subsection{ \nii\,/\ha\  in boxed regions}\label{box}
To determine the spatial variation (if any) of metallicity or ionization sources, we analyze the spatial distribution of  \nii\,/\ha\ ratios.
In order to enhance the S/N of the N2 line ratio map, we bin the pixels in five rectangular regions along the major axis and extract spectra for each boxed region (Figure~\ref{fig:ha}). 
The widths of the regions are chosen in such a way that each region contains sufficient  pixels to extract a reliable spectrum (i.e., S/N $>$ 3 for the \nii\ line). 
We derive the spectrum for each region using the \ha\ flux weighted median of all pixels in that region.  We fit the \nii\ and \ha\ emission lines and obtain the 
errors on \nii\,/ \ha\ by propagating the flux errors of \nii\ and \ha.

The peak of  the \ha\ emission corresponds to the center of Box 3 in Figure~\ref{fig:ha}.   We do not find significantly enhanced \nii\,/\ha\ ratios in the central pixels and rule out 
 AGN contamination of this target.

Our \nii\,/\ha\  emission-line ratios are shown in Figure~\ref{fig:box}.  The  \nii\,/\ha\  ratios of Box 3 and 4 are 0.303$\pm$0.034 and 0.312$\pm$0.033 respectively, close to solar values.  The  \nii\,/\ha\  ratios for Box 1 and 2 are 0.472$\pm$0.046 and 0.384$\pm$0.03 respectively,  significantly higher than those of  Box 3 and Box 4.  
We can only derive an upper limit of \nii\,/\ha\  ($<$0.233) for Box 5.  The high \nii\,/\ha\ values in the non-central regions of the galaxy indicate either extremely high metallicity
or the presence of  non-photoionization excitation.   

We further analyze individual pixels in Box 1 and 2 and find that the  \nii\,/\ha\ ratio is the highest in
 a sub-region highlighted as Box B in Figure~\ref{fig:ha}.  Very interestingly,  Box B which has the highest \nii\,/\ha\  ratio  (\nii\,/\ha\ =0.499$\pm$0.05) has also the ``bluest" velocity ($\Delta {\rm v} \sim 150 \kms$) with respect to the center of Box 3.  Unfortunately, the data quality is not sufficient to reveal any significant
 variations in the velocity dispersion map.   As discussed in the previous section, large 
 \nii\,/\ha\ ratios may indicate non-photoionized emission and therefore we do not convert this ratio to metallicity.
We note that the consequences of deriving metallically using  \nii\,/\ha\  ratios  in this case will result in the conclusion of a flat or inverted metallicity gradient (Appendix). 
In the next section, we discuss the possible origins of these elevated \nii\,/\ha\ ratios.

\section{Discussion}
\begin{figure}[!ht]
\hspace{-0.4cm}
\includegraphics[scale=0.46,angle=90]{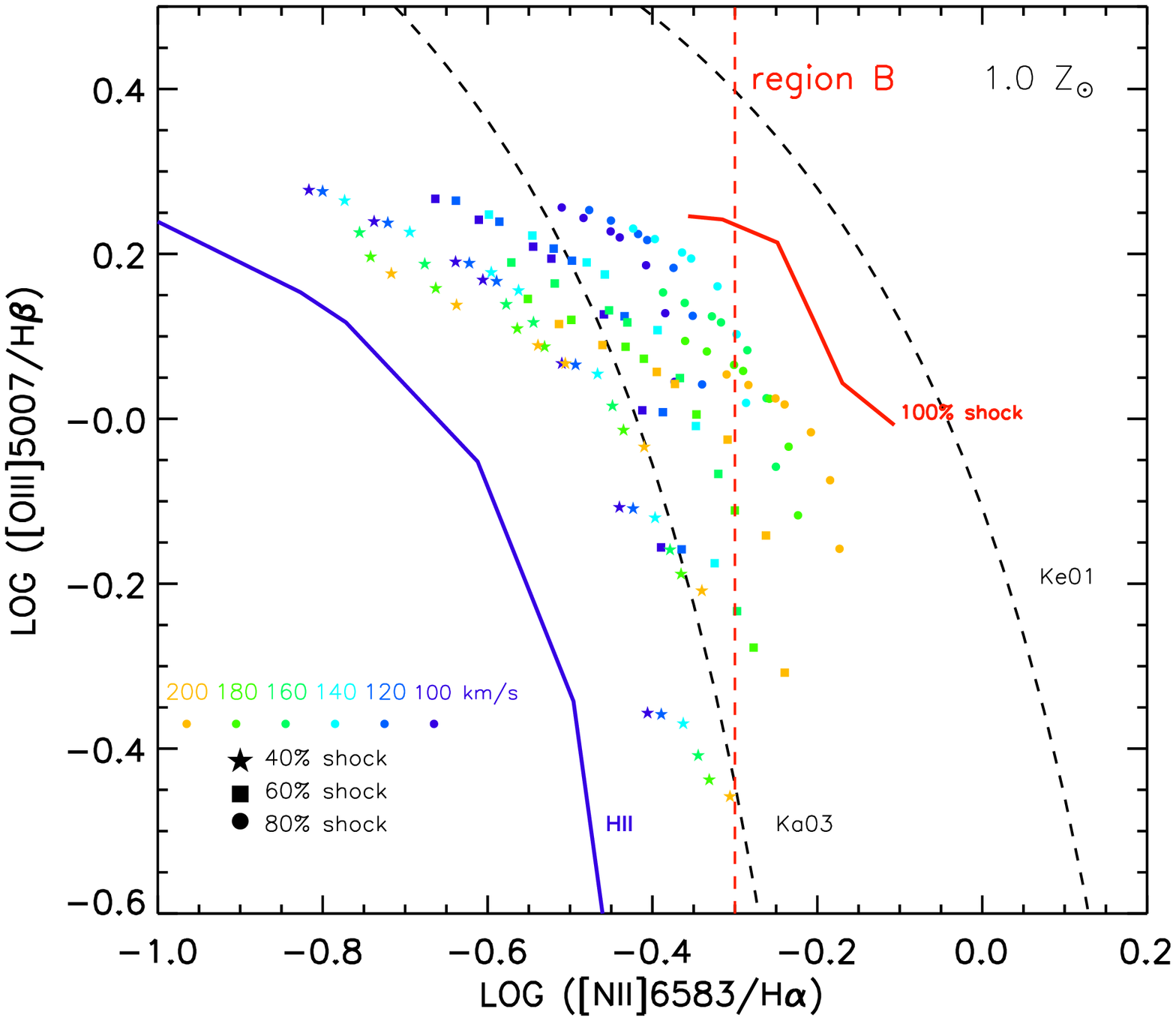}
\caption{The starburst-shock mixing grids on the \oiii\,/\hb\ versus \nii\,/\ha\ diagram. 
The dashed curve labeled as ``Ke01" is  the theoretical  ``maximum starburst line" derived by \citet{Kewley01b} as an upper envelope for star-forming galaxies.
The dashed curve labeled as ``Ka03" is the \citet{Kauffmann03} semi-empirical boundary for star-forming galaxies.   
The pure HII region models are shown as the blue solid line and the pure shock models are shown as the red solid line. 
The colored dots in between form a mixing sequence from pure HII region to pure shock excitation. Different colors represent different shock velocities and 
the relative contribution of shock is shown in different symbols. Solar abundance is adopted. The \nii\,/\ha\ ratio for region B in Figure~\ref{fig:box} is highlighted 
as  the vertical red dashed line.}
\label{fig:shock}
\end{figure} 

The standard BPT diagrams prove to be a powerful tool for investigating the ionization
sources in galaxies \citep{Baldwin81,Veilleux87,Kewley01b,Kauffmann03,Kewley06}. Employing IFS data on the BPT diagram has revolutionized our 
understanding of the power sources across entire galaxies \citep[e.g.,][]{Rich10,Rich11,Rich12,Sharp10}.  
The  \nii\,/\ha\  versus \oiii\,/\hb\ diagram is useful for distinguishing pure star-forming regions from non-photoionized  regions. 
The \nii\,/\ha\ ratio increases linearly with nebular metallicity until high metallicities where the log(\nii\,/\ha) ratio saturates at -0.5 \citep{Denicolo02,Kewley02,Pettini04}.
Any non-thermal contribution will shift the \nii\,/\ha\ ratio towards higher values than this saturation level \citep{Kewley06,Rich10,Rich11}. 
The  log(\nii\,/\ha) ratio of region B is 0.2 dex larger than the  saturation level of $-0.5$, suggestive of shock ionization.

We apply a mixed grid of photoionization and shock models to quantify the potential relative contribution of shock and \hii\ region ionization to the  \nii\,/\ha\ ratio. 
 The \hii\ region models are taken from the most recent Starburst99(SB99)/Mappings III model suite \citep{Leitherer99, Sutherland93}.  The model setup and input parameters are similar to those of \citet{Levesque10a} except that we use a slightly higher mass cut of 120 M$_{\odot}$ for the Salpeter initial mass function (Salpeter 1955).   We use a continuous star formation history with age 5.0 Myr and $n_e$=100~cm$^{-3}$.  Solar metallicity was used with a range of ionization parameters log $\mathcal{U}$ ($-3.5$ to $-1.9$).   We then mix the \hii\ region model with the slow-velocity shock model as described in \citet{Sutherland93,Farage10,Rich10,Rich11}.  We assume a 100\% pre-ionization, with a slow shock velocity ranging from 100 to 200 km s$^{-1}$, in increments 20 km s$^{-1}$.    These shock
 velocities are typical of galactic winds \citep{Shopbell98,Rich10,Rich11,Sharp10}.

Our shock and starburst models are shown in Figure~\ref{fig:shock}.  We show mixing lines to represent the relative contributions of starbursts and shocks from 0 to 100\%. 
 Shocks are able to produce higher \nii/\ha\ ratios than pure starburst models for a given metallicity.  According to our models, a 20\% shock contribution could increase the  log(\nii/\ha) value 
 by 0.2~dex. The region B of the A2667arc has log(\nii/\ha)=$-0.3$, lying well within the region  covered by the shock models.

 Shocks induced by galactic winds/outflows associated  with either stellar winds, supernovae or nuclear activity are commonly observed in local galaxies \citep[e.g.,][]{Veilleux05,Rupke05,Rich10,Soto12}.  The ubiquity of winds is even more prominent at high redshift given their violent star formation activities \citep[e.g.,][]{Pettini01,Swinbank05,SatoT09,Weiner09,MartinCL09,MartinCL12,Rubin10,Rubin11,Alexander10,Steidel10}.    In the cold flow scenario, the turbulent disks are associated with shocks \citep{Genzel11}.    The coincidence of region B being  blue-shifted with respect to the center strengthens the argument that it is an outflow shocked region.
\begin{figure}[!ht]
\begin{center}
\includegraphics[scale=0.36,angle=90]{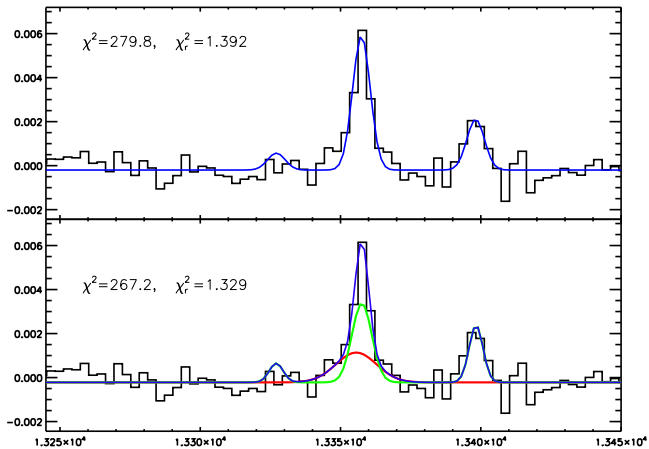}
\caption{The co-added spectrum of Boxes 1, 2, 4, 5 (i.e., non-nuclear regions) in Figure 6. 
The x axis is  wavelengths  in \AA, and the y axes are  flux density are in units of 10$^{-17}$ ergs~s$^{-1}$~cm$^{-1}$~\AA$^{-1}$.
The top panel fits a single component to the \ha\ and \nii\ lines (blue solid curve). 
The bottom panel fits a narrow  component (green curve) to the \ha, \nii\ lines and adds
a broad component (red curve) to the \ha\ line; the sum of the two-component fit is shown (blue curves).
The broad \ha\ component on the bottom panel (red line) is   $\sim$ 60 km~s$^{-1}$ blue-shifted from the narrow-line component. }
\label{fig:fitbroad}
\end{center}
\end{figure} 

Indeed, the outflow scenario is further supported by the detection of  a broad component in \ha. 
We detected a broad component in the co-added spectrum of boxes 1, 2, 4, 5.
In Figure~\ref{fig:fitbroad} we show that by including a broad  component in \ha, the fitting to the co-add spectrum is improved 
by 3$\sigma$  in a $\chi^{2}$ sense.  The broad component indicates a broadening wing of 
$\Delta \rm{v} (FWHM)\sim380\pm130$ km~s$^{-1}$, with a slight blueshift 
of $\sim$ 60 km~s$^{-1}$ with respect to the narrow line component.  The contribution of the broad wing to the total \ha\ flux of the co-added spectrum
is $f_{broad}\sim$ 40\%.   

Because of the unknown orientation of the A2667arc system, the outflow velocity is difficult to estimate. We thus compare the 
outflow properties of the A2667arc with other studies using the observed linewidth $\Delta \rm{v} (FWHM)$.  For example,
NGC 1569,  a local dwarf starburst galaxy, has a wind-excited broad component of $\Delta v (FWHM) \leq 300$~km~s$^{-1}$ \citep{Westmoquette07}. 
Arp 220, a famous local IR-luminous starburst galaxy, has a nuclear wind broadening of $\Delta v (FWHM) \sim 300-800$~km~s$^{-1}$ \citep{Arribas01}.  For a large sample of 39 local   ULRGs studied in \citet{Soto12b}, the median line-width for the shocked components  is  $\Delta v (FWHM) \sim 339 $~km~s$^{-1}$ (after converting  their 
 $\sigma$ to FWHM using  FWHM= 2.3548$\times$Gaussian$\sigma$). The median SFR of the \citet{Soto12b} ULIRGs sample is $\sim$ 73  M$_{\odot}$ yr$^{-1}$.
At very high redshift ($z >$ 2), the UV-selected massive SINS galaxies are found to have $\Delta v (FWHM) \sim 500$~km~$s^{-1}$ \citep{Shapiro09,Genzel11},  with  a median SFR of  $\sim$ 72  M$_{\odot}$ yr$^{-1}$.  Roughly, the broadening linewidth ($\Delta \rm{v} (FWHM)\sim380\pm130$ km~s$^{-1}$) and SFR (49.2$\pm$16.5 M$_{\odot}$ yr$^{-1}$) of A2667arc is most similar to the local ULIRGs sample of  \citet{Soto12b}. However, the similarity between the linewidths and SFR of the A2667arc and the $z\sim2$ SINS galaxies cannot be ruled out within 3$\sigma$ of the $\Delta \rm{v}$ (FWHM) measurement uncertainty.


If, on the other hand, the high extra-nuclear \nii\,/\ha\ ratio regions in the A2667arc are caused by extremely high metallicity,  
we lack theories to explain how  outer regions with super-solar metallicity would evolve spatially as a function of time. 
It is unclear how the A2667arc has gone through such a quick enrichment. Throughout the study we use the peak of \ha\ as the ``galactic center", which does not necessarily concur with the true galactic center.  However, the choice of center would not change our main conclusion that there is a shocked region blue-shifted from the peak of \ha\ emission.

Finally, the possibility of shocks will cause  an overestimation of the ``true" SFR. 
The \ha\ flux in  Regions 1 and 2 is $\sim$20\% of the total \ha\ flux.
\citet{Rich10} showed that for a slow shock of $\sim$ 100$\kms$, 50\% to 70\% of the \ha\ luminosity can be ascribed to shocks. Hence, $\sim$10\% of the total \ha\
flux of A2667arc may result from shocks.

\section{Summary}
We present the OSIRIS IFU data analysis of the spatially resolved giant arc at z\,=\,1.03 in Abell 2667.  
With a flux magnification of 14.0$\pm$2 in flux and $\sim 2 \times 7$ in area, we achieved a spatial resolution of 115$-$405 pc on the source plane. 
We find the A2667arc is clumpy in rest-frame UV and optical morphology, and has a high velocity dispersion  V$_{rot}$ ${\it \sin{i}}$/$\sigma\sim$ 0.73, 
similar to the turbulent star forming disks at z$\sim$ 2.  The current morphology and velocity map is not sufficient to distinguish whether the A2667 arc is 
a face-on clumpy disk or a merger.   We find that  \nii\ and \ha\ ratios first rise and then fall when moving out from the galactic center
defined as the peak of \ha\ flux.  The central 350 pc of the \ha\ nucleus region is dominated by star-formation with metallicity \oh\,$=$8.57$\pm$0.03 ($\sim$ 0.76 $Z_{\odot}$). 
The \nii\,/\ha\ ratios in the 350 $-$ 900 pc annulus show a significantly enhanced value ($\sim$ 0.1 dex higher than the nucleus). The highest \nii\,/\ha\ ratio region ($\sim$ 0.2 dex higher) corresponds to a $\sim$100 $\kms$  blue-shifted velocity clump with respect to the nucleus.  Applying the most recent slow-shock models, we propose that this elevated \nii/\ha\ ratio is caused by shock excitation from a galactic 
wind. Our results indicate that extreme caution must be used when interpreting line ratios as metallicity gradients at high redshift. Shocked extra-nuclear regions can mimic flat or inverted 
metallicity gradients.  

\acknowledgments 
This work is based on data obtained at the W. M. Keck
 Observatory. We are grateful to the Keck Observatory staff for assistance with our observations, especially Jim Lyke, and 
Randy Campbell.   T.Y. thanks the hospitality of the Australian National University and a Founder Region Fellowship for Women. 
L.K. acknowledges a NSF Early CAREER Award AST 0748559 and an  ARC Future Fellowship award FT110101052. 
 A.M.S. acknowledges a STFC Advanced Fellowship.  JR is supported by the Marie Curie Career Integration Grant 294074. 
The authors recognize the very significant cultural role that the summit of Mauna Kea has 
 within the indigenous Hawaiian community.  


\begin{table*}
\begin{center}
{\footnotesize
{\centerline{\sc Table 1.}}
{\centerline{\sc Derived Global Physical Properties for Arc A}}
\begin{tabular}{ccccccccc}
\hline
\noalign{\smallskip}

Re    & $E\rm{(B-V)}$ &     SFR$_{\rm{H\alpha}}$     &$M_{\rm{star(SED)}}$         &    $\sigma$ & $V_{\rm{max}} \sin{i}$ & \oh& $\mu$, $\mu1$, $\mu2$ \\
 kpc  & SED fit    & $\Msolyr$               &$10^{10} \Msol$     & $\kms$ & $\kms$   &  Integrated & flux, axis1, axis2\\
\noalign{\smallskip}
\hline
\hline
\noalign{\smallskip}
\noalign{\smallskip}
  1.1$\pm$0.2  & 0.6    &  49.2$\pm$16.5   &1.90$^{+2.0}_{-1.0}$           & 65$\pm$6 &  48$\pm$12 & 8.606$\pm$0.051 & 14$\pm$2.1, 2.0$\pm$0.4,7.0$\pm$1.4 \\
\noalign{\smallskip}
\hline
\label{tab:sum}
\end{tabular}
}
\tablecomments{Derived intrinsic global properties for A2667 arc A. Col. 1: The half light radius from fitting a Sersic profile to the red $\emph{HST}$ component. 
Col. 2: Extinction derived from SED fitting. Col. 3: SFR from \ha\ flux, corrected for lensing magnification and extinction. The error comes from the uncertainty
in magnification and flux calibration.
Col. 4: Stellar mass from SED fitting.  Col. 5:   Median velocity dispersion, weighted by \ha\ flux.  Col. 6: $V_{max} \sin{i}$ see the text for methodology. Col. 7:  Metallicity from  the integrated spectrum. Col. 8: Flux magnification and size magnification  
in the major and minor axes.}
\end{center}
\end{table*}


\appendix

\section{\nii\,/\ha\ as a function of radius}
\begin{figure*}[!ht]
\hspace{-0.2cm}
\begin{center}
\includegraphics[scale=.8,angle=0]{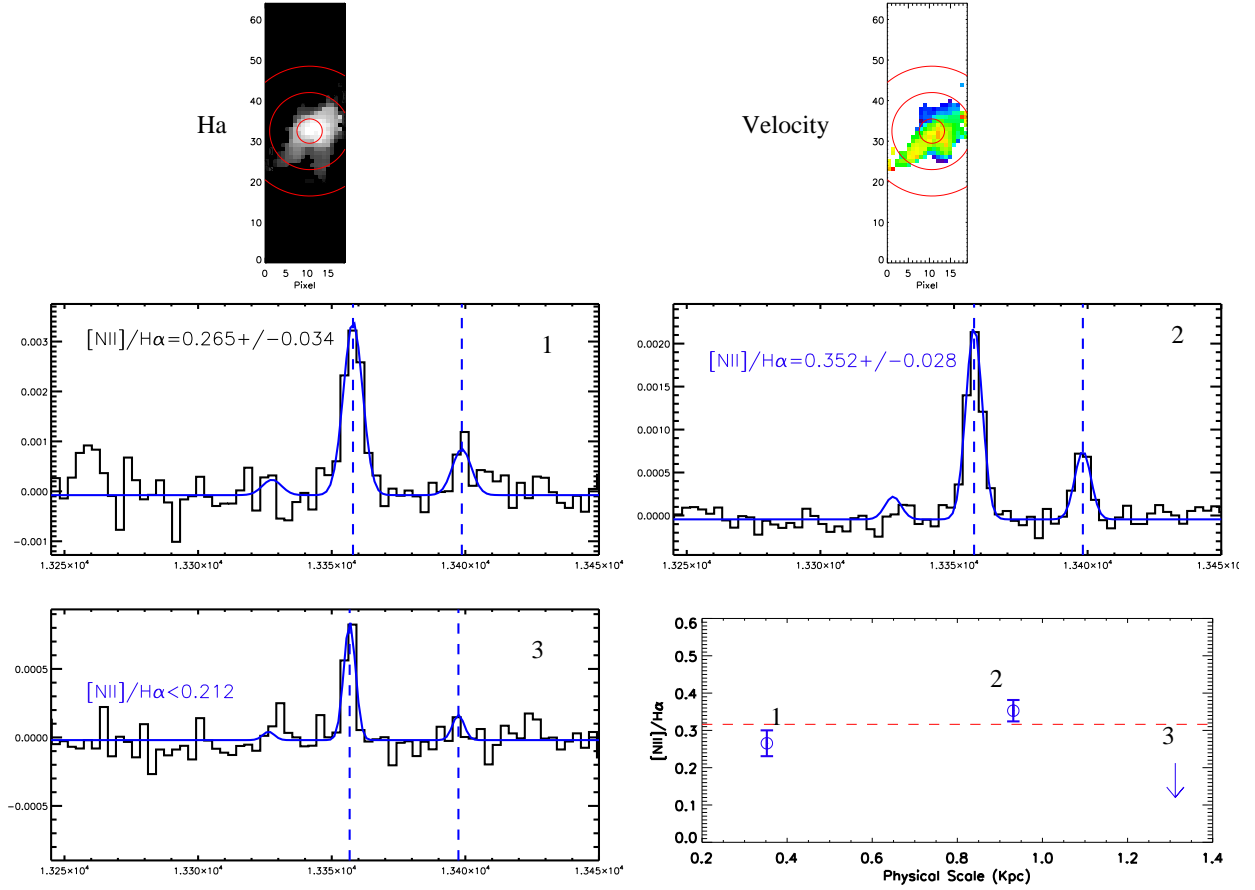}
\caption{The \nii\,/\ha\ ratio as a function of the ``galactocentric" distance. The galactic center is defined as the peak of \ha. 
 The top left panel shows the 3 annuli relative to the \ha\ intensity map. The top right panel 
shows the 3 annuli relative to the velocity map. Note that these images are given in IFS datacube coordinates, thus North and East is 37\degree\ clockwise from the up and left position.  The following 3 panels show the 
 observed image-plane spectra  averaged in the 3 annuli. The wavelength is in units of  \AA, and flux density is in units of  10$^{-17}$ ergs~s$^{-1}$~cm$^{-1}$~\AA$^{-1}$.   The line positions are marked on the plot . The last panel shows the \nii\,/\ha\ ratio as a function of the intrinsic  ``galactocentric" distance (after lensing correction).  The red horizontal line indicates the 
line ratio corresponding to the solar \nii\,/\ha\ ratio. Ratios above solar may be contaminated by non-photoionized emission and can not be directly related to metallicity.  
}
\label{fig:rad}
\end{center}
\end{figure*}

In metallicity gradient studies, averaging spectra in annular regions is a common exercise \citep{Moustakas10, Queyrel11, Yuan11}. 
To relate to these studies, we extract spectra in 3 annuli centered on the \ha\ peak (Figure~\ref{fig:rad}).  Each spectrum is  calculated as the  \ha\ flux weighted median of all pixels within the annulus. 
From Section~\ref{box}, we see that the line ratio map is not symmetric with respect to the center. Averaging spectra in annuli smooths out the lateral spatial variance, resulting in noisier \nii\ lines.   The choice of the annuli is such that the integrated spectra gives S/N $>$ 5 for the \nii\ lines. 
The physical lengths of the three outer radii of the annuli in the source plane are  0.35, 0.93, and 1.31 kpc respectively. 
The \nii\ line is robustly detected at S/N $>$ 5 for the inner two annuli and is a 3$\sigma$ detection for the outer annulus. 
 
The  \nii\,/\ha\ ratios are larger in the circum-nuclear region (350$-$930 pc) than in the nucleus ($<$350pc).
Small \nii\,/\ha\ ratios are seen in the outer annulus (930$-$1.3 kpc). If  the \nii\,/\ha\ ratios arise purely from \hii\ regions, these
\nii\,/\ha\ ratios could translate into a  flat  or inverted metallicity gradient similar to those in several  high redshift studies such as \citet{Cresci10, Queyrel11}. 

However, in A2667arc, these enhanced circum-nuclear \nii\,/\ha\ ratios are remarkably similar to the LINER-like luminous infrared galaxies reported in \citet{Rich10,Rich11}.
Rich et al. observed that enhanced extra-nuclear \nii\,/\ha\ ratios in late-stage galaxy mergers are driven by slow shock emission (100$-$200 km~s$^{-1}$) associated
with galactic-wide winds or gas collisions due to mergers.  Rich et al. (2010, 2011) showed that both velocity dispersions and the strong line ratios independently indicate shock excitation 
in the extra-nuclear regions. Slow shock models confirm this conclusion, yielding remarkable agreement with the enhanced emission-like ratios.

 \end{document}